# Measuring the Localization Length through the superconductor-insulator transition in ultrathin amorphous beryllium films

Wenhao Wu[+] and E. Bielejec[*]

[+]Department of Physics, Texas A&M University, College Station, Texas 77843

[*]Sandia National Laboratories, Albuquerque, New Mexico 87185

**Abstract**

Electron transport and tunneling across the superconductor-insulator (SI) transition have been measured simultaneously for quench-condensed ultrathin amorphous beryllium films. The anomalous negative magnetoresistance previously observed in insulating films disappears when Mn impurities are introduced to the films, restoring a rather clean Efros-Shklovskii type hopping behavior. The combination of transport and tunneling data allows us to determine, *independently* and up to a constant on the order of unity, the localization length, $\xi_L$, and the dielectric constant, $\kappa$, for the films. As the normal-state sheet resistance of the films at 20 K is reduced with increasing film thickness, $\xi_L$ increases exponentially. The SI transition occurs when $\xi_L$ crosses the Ginzburg-Landau coherence length, $\xi_S$.

PACS Numbers: 74.25.Dw, 73.20.Fz, 71.30.+h, 72.20.Ee

Superconductivity in the presence of disorder has been a subject of great interest over the years [1]. In particular, recent discoveries [2,3] of the superconductor-insulator (SI) transitions in two-dimensions (2D), tuned by disorder or magnetic field are fascinating [4,5]. Theoretical studies of the effects of disorder on superconductivity date back to Anderson's theorem [6] which predicts that nonmagnetic impurities have no significant effect on superconductivity. However, this theorem does not consider the effect of Anderson localization. The scaling theory of localization [7] predicts that, in zero magnetic field, electronic wave functions are always localized in disordered 2D systems over a length scale called the localization length, $\xi_L$. Since superconductivity is a manifestation of long-range phase coherence of paired-electron states, the fundamental question is when and how superconductivity breaks down as the electronic states become strongly localized with increasing disorder.

A number of theories [8,9] have addressed this question. With increasing disorder, one expects that, when $\xi_L$ becomes smaller than the size of a vortex core, the vortex core becomes an insulator. The vortex core could disintegrate because of large fluctuations in the phase of the superconducting order parameter. Robust supercondcutivity exists when $\xi_L > \xi_S$, where the Ginzburg-Landau coherence length, $\xi_S$, is a measure of the size of the vortex core. The other criterion [9] to consider is $[N(0)\xi_L^d]^{-1} < \Delta_S$, where $N(0)$ is the average density of states (DOS) at the Fermi energy, $d = 2$ is dimensionality, $[N(0)\xi_L^d]^{-1} = \Delta_L$ is the typical energy separation between localized states, and $\Delta_S$ is the Bardeen-Cooper-Schrieffer (BCS) gap. This condition means, if a localized system is divided into boxes of size $\xi_L$, there are on average a few localized states in each box with energies within $\Delta_S$ of the Fermi surface. These states can form a coherent local superconducting fluctuation. A superconducting ground state can be stabilized via inter-box Josephson tunneling established by localized states connecting adjacent boxes [10].

Recent studies of the SI transition have largely been motivated by experiments on ultrathin disordered superconducting films [2,3,11-14]. Many of these studies analyze the data following a scaling argument based on a Bose-Hubbard model [4] in which disorder drives a Bose-condensed superfluid into an insulating glass. However, this scaling theory does not address the microscopic origin of the SI transition. In this Letter, we investigate the interplay of localization and superconductivity in quench-condensed ultrathin amorphous Be films of about 10 Å in thickness. We directly probe one of the most fundamental length scale of a disorder system, the localization length $\xi_L$, across the SI transition. $\xi_L$ has been studied in a number of systems [15] by fitting the conductance to certain hopping laws. However, these experiments *do not* measure $\xi_L$ directly. Rather, they measure the product $\xi_L\kappa$, where $\kappa$ is the dielectric constant. In general, $\kappa$ is not a constant and can change drastically near a phase transition. Here, we describe a procedure with which $\xi_L$ and $\kappa$ can be determined *independently* across the SI transition, based on simultaneous transport and tunneling measurements.

The Be films used for this study were thermally evaporated onto glass substrates held near 20 K. A multi-lead pattern with an area of 3×3 mm$^2$ between neighboring leads was pre-evaporated on the substrates. Details of our transport and tunneling measurements have been published earlier [14,16,17]. Figure 1 shows the sheet resistance, $R_\square$, versus $1/T^{1/2}$ on a semi-log plot for a set of Be films following an evaporation sequence to increase film thickness. Data in Fig. 1 show different behavior in two temperature regimes. At high temperatures, above 5 K, the curves appear to follow straight lines converging to about 10 kΩ/□ in the T→ ∞ limit. Below 5 K, $R_\square$ increases much steeper with decreasing T for insulating films. These insulating films also show a puzzling mangetoresistance as we [16] have reported: the resistance can *decrease* by orders of magnitude with an increasing perpendicular field in the range of 2-10 T. Similar behavior was later observed in InOx films and was interpreted [18] as the field suppression of the super-insulating behavior due to a nonuniform $\Delta_S$ which leads to regions of large $\Delta_S$ interceded with regions of vanishingly small $\Delta_S$, in analog to granular superconducting films. Since our magnetic field is not strong enough to completely suppress this super-insulating behavior in the Be films, we have recently introduced a small amount of Mn impurities to investigate the effect of magnetic impurities on the magnetoresistance. Mn was loaded in a tungsten basket, with the evaporation rate-current relation calibrated in a thermal evaporator at room temperature. The Mn source was mounted together with a Be source in a multi-source quench-condensation cell at the bottom of the IVC can of a dilution refrigerator. A mechanical shutter inside the dilution refrigerator, manually controlled from the top of the refrigerator, was used to open and close the Mn source during evaporation. The actual amount of Mn impurities deposited on the Be films was too small to be registered by the quartz thickness monitor mounted inside the dilution refrigerator. In Fig. 2, we plot $R_\square$ versus $1/T^{1/2}$ measured on two Be film sections 2 mm apart, following a sequence of evaporations to increase the thickness of the film. The two curves at the top of Fig 2(a) were measured for two very thin film thicknesses (5-6 Å) before Mn was evaporated. These samples displayed the typical upward curvature for insulating films as seen in Fig 1 as well as the previously reported large and negative mangetoresistance [16]. We then evaporated a small amount of Mn impurities onto the Be film by manually opening the shutter briefly. We observe in Fig. 2(a) that the upturn in the $R_\square$, versus $1/T^{1/2}$ curve disappeared after Mn impurities were introduced (Curve 3 from the top), recovering a rather clean Efros-Shklovskii hopping behavior. In addition, the large negative magnetoresistance observed earlier [16] also disappeared. Such behavior persists for films produced following subsequent Be evaporation steps to increase the thickness, as shown by the remaining curves in Fig. 2. We believe this is strong evidence that the large and negative magnetoresistance observed in the absence of Mn impurities is due to the field-suppression of superconductivity [18].

The high temperature data in Fig. 1 and the data with Mn impurities in Fig. 2 converge to a unique resistance pre-factor of $R_0 \approx 10$ k$\Omega$/$\square$ in the $T \rightarrow \infty$ limit, which is close to a pre-factor of $R_0 \sim R_Q/2 = (h/e^2)/2 \approx 13$ k$\Omega$/$\square$ for the ES hopping law $R_\square(T) = R_0\exp(T_0/T)^{1/2}$ with $T_0 = 2.8e^2/(4\pi\varepsilon_0 k_B \xi_L \kappa)$, where e is electron charge, $\varepsilon_0$ is the vacuum permitivity, and $k_B$ is the Boltzmann constant. This hopping law is predicted for the ES soft Coulomb gap [19] in the density of states (DOS). We fit all our data to the ES hopping law to obtain parameters $T_0$ and $R_0$. We use the sheet resistance of the films at 20 K, R, as the disorder parameter. The convergence of the curves seen in Fig. 1 is significant in that it is observed for all the films that are either insulating or in the vicinity of the SI transition, with R varying by nearly three orders of magnitude as shown in the lower frame of Fig. 2. Using $T_0$, we obtain the product, $\xi_L\kappa = 2.8e^2/4\pi\varepsilon_0 k_B T_0$. In Fig. 3, $\xi_L\kappa$ and $R_0$ are plotted versus R.

Next, we use the measured tunneling DOS, shown in Fig 4 for one of the samples in Fig. 2, to obtain $\kappa$, so that $\xi_L$ can be extracted from the product $\xi_L\kappa$. The predicted [19] linear DOS is given by $N(E) = [\alpha(4\pi\varepsilon_0\kappa)^2/e^4]|E|$ with a slope of $\alpha(4\pi\varepsilon_0\kappa)^2/e^4$, where $\alpha$ is on the order of unity and E is energy measured from the Fermi energy. In practice, we use Al/Al$_2$O$_3$/Be junctions to measure the tunneling conductance G as a function of the bias voltage, V$_{bias}$. We then calculate the slope of the measured linear G-versus-V$_{bias}$ curve, which is proportional to $\kappa^2$ since G is proportional to the DOS. Using the slopes calculated from the tunneling conductance for samples of varying R, we have obtained a power-law dependence of $\kappa^2 \sim R^{-1.73 \pm 0.10}$, or, $\kappa \sim R^{-0.87}$. The absolute value of $\kappa$ is unknown, since we have not attempted to convert G to the DOS in real units. This conversion requires a calibration factor that is related to the product of the tunneling probability and the DOS of the Al counter-electrode. This calibration factor should be a constant for a given junction in the relevant temperature regime. We divide the product $\xi_L\kappa$ by $\kappa$ ($\sim R^{-0.87}$) to extract $\xi_L$. Since the absolute value of $\kappa$ is unknown, the value of $\xi_L$ so obtained also carries an unknown constant factor. However, it is reasonable to set $\xi_L$ to a value of $\xi_{L0}$ = 5 Å in the extreme insulating limit (R = 10$^3$ - 10$^4$ kΩ/□). We believe that this assumption is correct up to a numerical factor of 2 [20]. Following these steps, we have obtained $\xi_L$ as a function of the disorder parameter R, as shown in Fig. 5. With the absolute value of $\xi_L$ set, we divide the product $\xi_L\kappa$ by $\xi_L$ to set the absolute value for $\kappa$, which is plotted in the inset to the upper frame of Fig. 3. With this procedure, $\xi_L$ and $\kappa$ are determined, *independently*, with an uncertainty factor of about 2. The scaling theory of localization in 2D predicts in the weakly disordered limit that $\xi_L$ grows exponentially with decreasing disorder [1,7]: $\xi_L = l\exp[(\pi/2)k_Fl]$, where k$_F$ is the Fermi wave number and *l* is the elastic mean free path. In deriving this, the relation for conventional transport, R = (h/e$^2$)(1/k$_F$*l*) ≈ (26 kΩ)(1/k$_F$*l*), is used. This scaling theory is a perturbative expansion based on 1/k$_F$*l*. We note that, effectively, the value of k$_F$*l* crosses 1 for films of R ~ 26 kΩ/□. One may assume that not too far from k$_F$*l* = 1 the theory is still relevant to certain degree. Bearing this in mind and assuming *l* ≈ $\xi_{L0}$ for films of k$_F$*l* ≤ 1, we rewrite the expression for $\xi_L$: $\xi_L = l\exp[(\pi/2)(h/e^2)/R] \approx \xi_{L0}\exp[(\pi/2)(h/e^2)/R]$. We plot this relation as a solid curve in Fig. 5. The agreement between the data and the solid line is surprisingly good, as there is no adjustable parameter.

The SI transition in the Be films occurs at a critical resistance of R ≈ 11.7 kΩ/□, as indicated by the dashed line in Fig. 5. Now, we show that the transition occurs just as $\xi_L$ crosses $\xi_S$. We estimate the value of $\xi_S$ using the Ginzburg-Landau expression [21], $\xi_S$

= $0.855(\xi_0 l)^{1/2}$. The BCS coherence length is $\xi_0 = \hbar v_F/(\pi\Delta_0) = \hbar v_F/(1.76\pi k_B T_C) = 0.18\hbar v_F/(k_B T_C)$, where $\Delta_0$ is the BCS gap energy for T = 0 and $v_F$ is the Fermi velocity which is related to R by R = $(h/e^2)(1/k_F l)$. Based on the superconducting transition temperature $T_C \approx 7.0$ K for thick films of R ~ 2 k$\Omega$/□, $\xi_S$ is estimated to be 150 Å. Comparing this value of $\xi_S$ with $\xi_L$ in Fig. 5, we conclude that the SI transition occurs just as $\xi_L$ crosses $\xi_S$. We note that a direct comparison between $\Delta_L$ and $\Delta_S$ is difficult [20]. To estimate $\Delta_L$ near the SI transition, we use a free-electron model: $\Delta_L = [N(0)\xi_L^d]^{-1} = (a^2 E_F)/\xi_L^2$, where a ~ 1 Å is the atomic spacing, $E_F$ ~ $10^4$ K is the Fermi energy, and d = 2. This leads to a lower bound of 1 K for $\Delta_L$. The actual value for $\Delta_L$ should be higher since the DOS at the Fermi energy is significantly depressed from the free-electron result for films near the SI transition. Alternatively, one may argue that the characteristic energy of the insulating phase, $k_B T_0$, is in fact a reasonable measure of $\Delta_L$. In Fig. 5, we plot $T_0$ and $T_C$ for Be films of varying R. The $T_C$ values were obtained from two film sections of varying thickness, with the SI transition occurring at a critical resistance of 11.7 k$\Omega$/□. Here we observe that the SI transition occurs as $T_0$ becomes comparable to $T_C$.

In summary, we have determined, *independently* and up to a constant on the order of unity, the localization length, $\xi_L$, and the dielectric constant, $\kappa$, for quench-condensed ultrathin amorphous beryllium films. $\xi_L$ decreases exponentially with increasing disorder. Our data suggest that the SI transition occurs when $\xi_L$ crosses the Ginzburg-Landau coherence length, $\xi_S$. We gratefully acknowledge discussions with A. M. Goldman, Z. Ovadyahu, and I. Zharekeshev. This work was supported in part by NSF under Grant No. DMR-0551813.

# References

1. P. A. Lee and T. V. Ramakrishnan, Rev. Mod. Phys. **57**, 287 (1985); D. Belitz and T. R. Kirkpatrick, *ibid*. **66**, 281 (1994).

2. R. C. Dynes, J. P. Garno, and J. M. Rowell, Phys. Rev. Lett. **40**, 479 (1978); D. B. Haviland, Y. Liu, A. M. Goldman, *ibid*. **62**, 2180 (1989); H. M. Jaeger, D. B. Haviland, B. G. Orr, and A. M. Goldman, Phys. Rev. B **40**, 182 (1989); A. M. Goldman and N. Markovic, Phys. Today, **51**, No. 11, 39 (1998); J. M. Valles, Jr., R. C. Dynes, J. P. Garno, Phys. Rev. Lett. **69**, 3567 (1992); S.-Y. Hsu, J. A. Chervenak, and J. M. Valles, Jr., Phys. Rev. Lett. **75**, 132 (1995).

3. A. F. Hebard and M. A. Paalanen, Phys. Rev. Lett. **65**, 927 (1990).

4. M. P. A. Fisher, G. Grinstein, S. M. Girvin, Phys. Rev. Lett. **64**, 587 (1990); M. P. A. Fisher, *ibid*. **65**, 923 (1990).

5. S. Sondhi, S. Girvin, J. Carini, and D. Shahar, Rev. Mod. Phys. **69**, 315 (1997).

6. P. W. Anderson, J. Phys. Chem. Solids **11**, 26 (1959).

7. E. Abrahams, P. W. Anderson, D. C. Licciardello, and T. V. Ramakrishnan, Phys. Rev. Lett. **42**, 673 (1979).

8. A. Kapitulnik and G. Kotliar, Phys. Rev. Lett. **54**, 473 (1985); G. Kotliar and A. Kapitulnik, Phys. Rev. B **33**, 3146 (1986).

9. M. Ma and P. A. Lee, Phys. Rev. B **32**, 5658 (1985).

10. Y. Imry and M. Strongin, Phys. Rev. B **24**, 6353 (1981); Y. Shapira and G. Deutscher, *ibid*. **27**, 4463 (1983); A. Frydman and Z. Ovadyahu, *ibid*. **55**, 9047 (1997).

11. Y. Liu *et al*., Phys. Rev. Lett. **67**, 2068 (1991).

12. N. Mason and A. Kapitulnik, Phys. Rev. Lett. **82**, 5341 (1999).

13. N. Markovic *et al*., Phys. Rev. B **60**, 4320 (1999).

14. E. Bielejec and W. Wu, Phys. Rev. Lett. **88**, 206802 (2002).

15. F. W. Van Keuls *et al*., Phys. Rev. B **56**, 13263 (1997); M. Furlan, *ibid*. **57**, 14818 (1998); F. Hohls, U. Zeitler, and R. J. Haug, Phys. Rev. Lett. **88**, 036802 (2002).

16. E. Bielejec, J. Ruan, and W. Wu, Phys. Rev. B **63**, R100502 (2001).

17. E. Bielejec, J. Ruan, and W. Wu, Phys. Rev. Lett. **87**, 036801 (2001).

18. G. Sambundamurthy, L. W. Engel, A. Johansson, and D. Shahar, Phys. Rev. Lett. 92, 107005 (2004).

19. B. I. Shklovskii and A. L. Efros, *Electronic Properties of Doped Semiconductors* (Springer, New York, 1984).

20. D. Shahar and Z. Ovadyahu, Phys. Rev. B **46**, 10917 (1992).

21. M. Tinkham, *Introduction to Superconductivity* (New York, McGraw Hill, 1996).

**Figure Captions**

Fig. 1  Selected curves of $R_\square$ versus $1/T^{1/2}$ for one Be film section following deposition steps to increase film thickness (from top to bottom). The thickness for these films changed from 4.6 Å to about 10 Å. The straight lines are draw as a guide for eye, showing that in the high-T regime all the curves follow straight lines that converge to about 10 kΩ/□ in the T→ ∞ limit. The films for bottom curve is superconducting at low temperatures.

Fig. 2  Upper frame: $R_\square$ versus $1/T^{1/2}$ for one Be film section following a sequence of evaporation steps. After the second curve from the top were measured, Mn impurities were evaporated onto the film, which resulted in the third curve from the top. Other curves resulted from additional Be evaporation steps. Lower Frame: Same data from the frame plotted on a broader temperature range.

Fig. 3  The product $\xi_L\kappa$ and $R_0$ are plotted versus the disorder parameter R. Data points shown here were obtained on forty-one films of various thickness fabricated in four experimental runs. The dielectric constant κ, with its absolute value fixed as described in the main text, is plotted in the inset to the upper frame.

Fig. 4  Tunneling conductance versus the bias voltage measured at 3 K on one Be film section in Fig. 2, showing the linear DOS near the Fermi energy. Curves from the bottom to the top correspond to values of the disorder parameter R = 3740, 298, 119, 37.3, and 14.3 kΩ, respectively.

Fig. 5  The localization length, $\xi_L$, is plotted as a function of the disorder parameter R. The absolute value of $\xi_L$ is fixed at $\xi_{L0} = 5$ Å in the extreme insulating limit. The solid curve is for $\xi_L = \xi_{L0}\exp[(\pi/2)(h/e^2)/R]$. The dashed line indicates the critical resistance for the zero-field SI transition, separating the superconducting phase on the left from the insulating phase on the right.

Fig. 6  The parameter $T_0$ (solid circles) from fits to the ES hopping law and the measure superconducting transition temperature $T_C$ are plotted versus the disorder parameter R. Up and down triangles represent data from two different experimental runs. The critical sheet resistances is 11.7 k$\Omega$/□ as indicated by the dashed line.

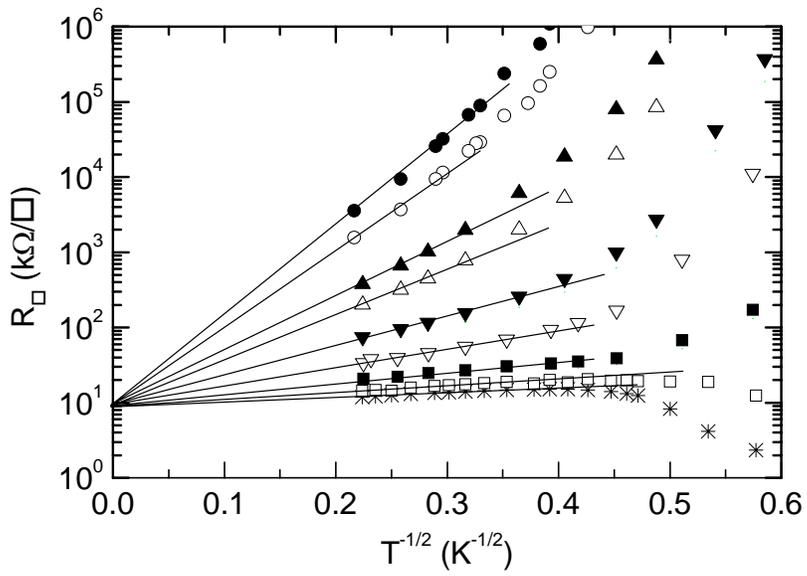

Figure 1  W. Wu et al.

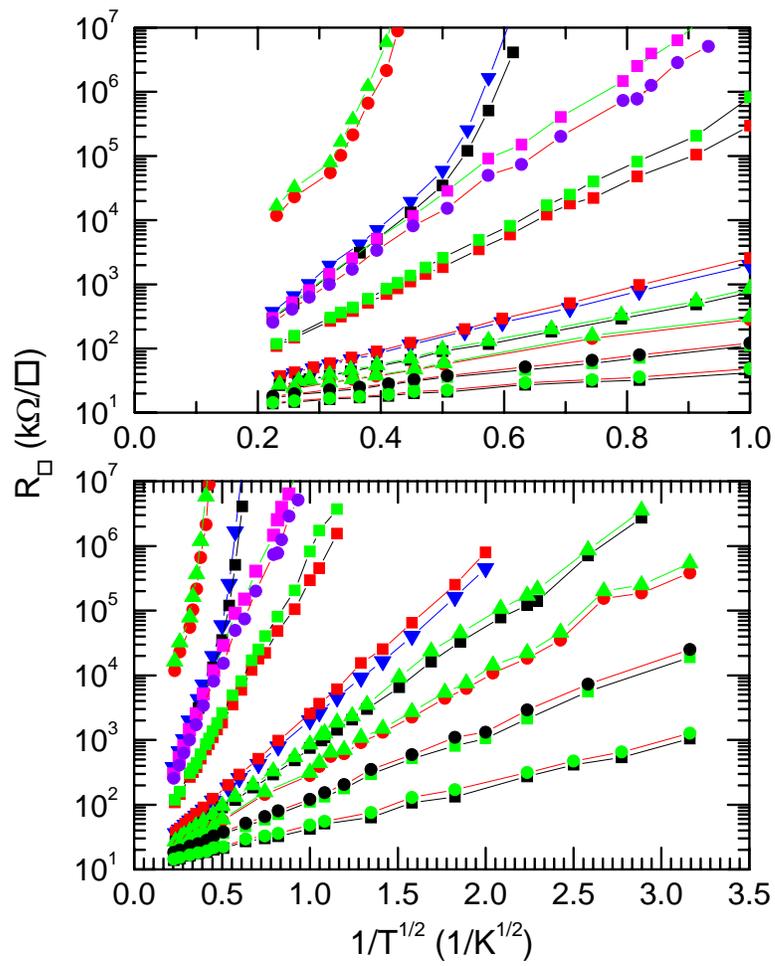

Fig. 2

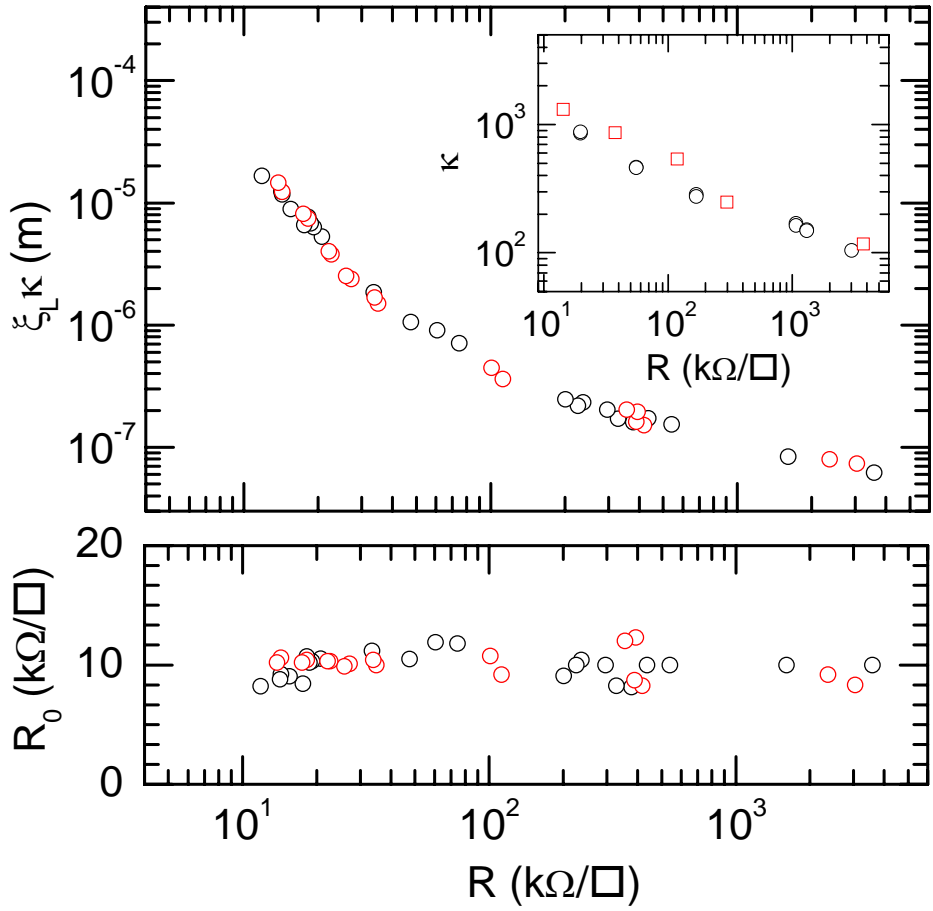

Figure 3 W. Wu et al.

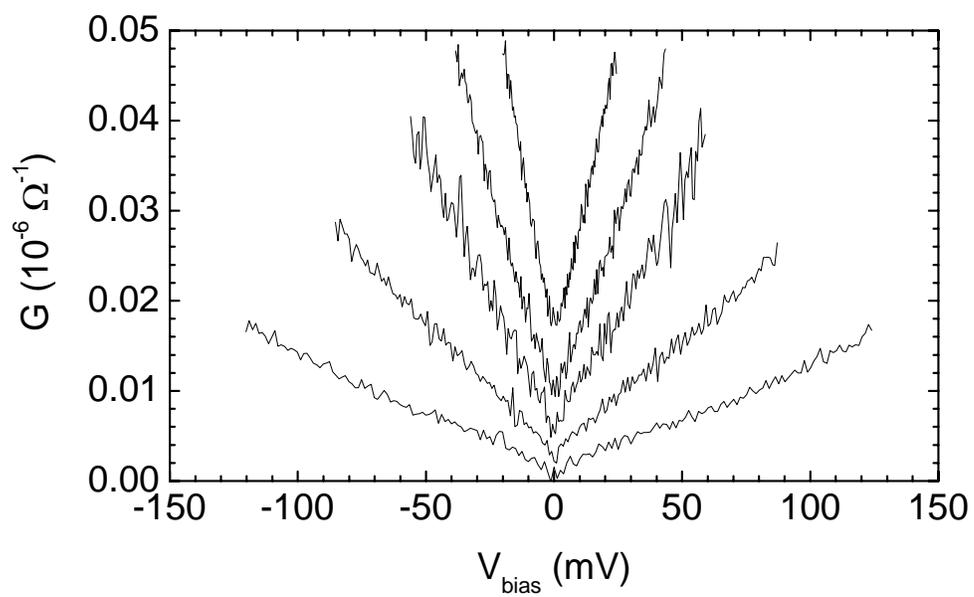

Fig. 4  W. Wu et al.

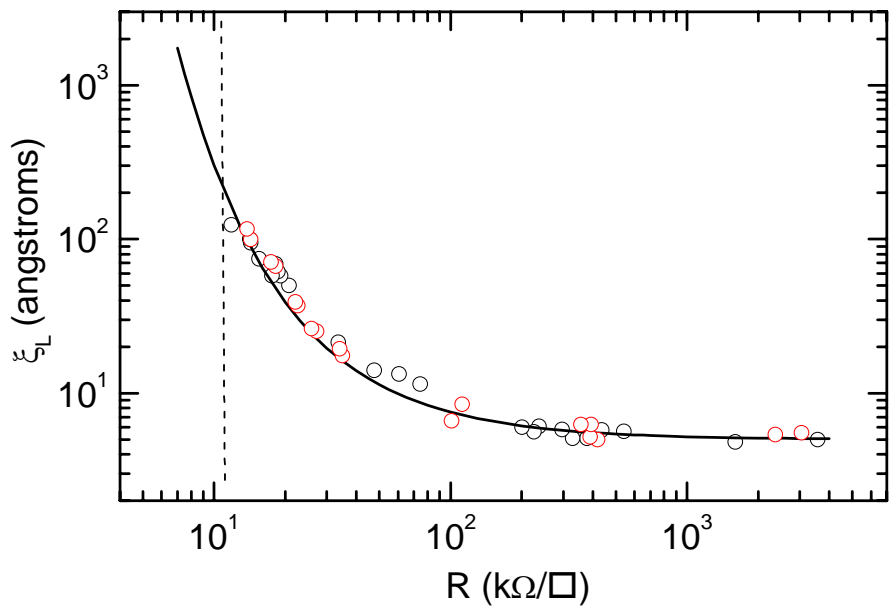

Figure 5    W. Wu et al.

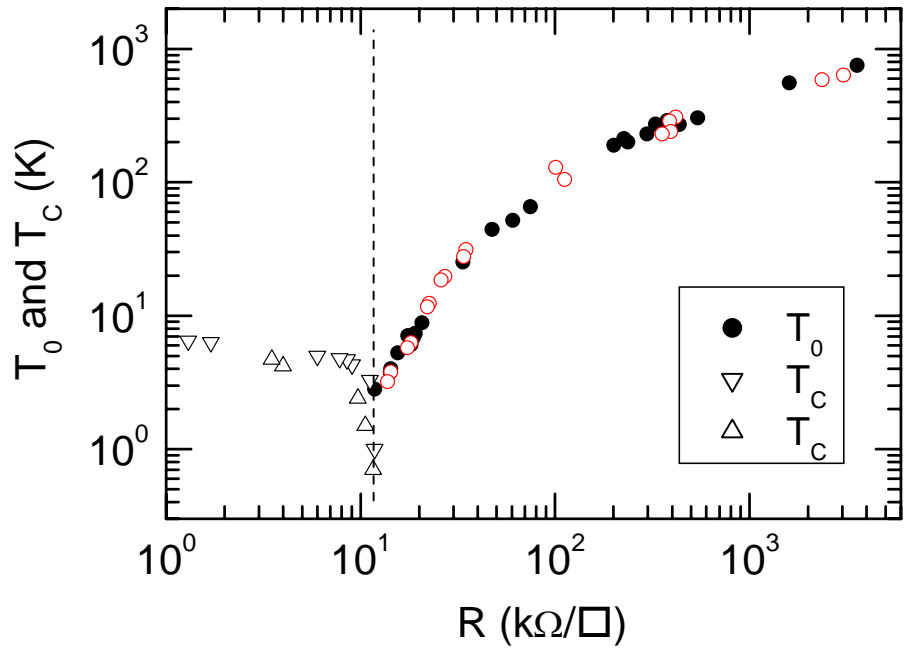

Figure 6     W. Wu et al.